\documentclass[5p,times,twocolumn]{elsarticle} 
\usepackage{amssymb, amsmath, amsbsy, wasysym}
\usepackage{algorithmic, algorithm}
\usepackage{algorithm}
\usepackage{listings}
\usepackage[T1]{fontenc}
\usepackage{color}
\journal{Computer Physics Communications}

\newcommand{\clear}[1]{ }

\begin{document}
\begin{frontmatter}

\title{Memory-efficient Lattice Boltzmann Method for low Reynolds number flows}
\author[ift]{Maciej Matyka\corref{cor}}
\ead{maciej.matyka@uwr.edu.pl}
\cortext[cor]{Corresponding author}
\address[ift]{Faculty of Physics and Astronomy, University of Wroc{\l}aw, pl.\ M.\ Borna 9, 50-205 Wroc{\l}aw, Poland, tel.: +48713759357, fax: +48713217682}
\author[icm]{{Micha{\l} Dzikowski}}
\address[icm]{{Interdisciplinary Centre for Mathematical and Computational Modelling, UW, ul. Tyniecka 15/17, 02-630, Warsaw, Poland}}

\begin{abstract}
The Lattice Boltzmann Method algorithm is simplified by assuming constant numerical viscosity (the relaxation time is fixed at $\tau=1$). This leads to the removal of the distribution function from the computer memory.
To test the solver the Poiseuille and Driven Cavity flows are simulated and analyzed. The error of the solution decreases with the grid size L as $L^{-2}$. Compared to the standard algorithm, the presented formulation is simpler and shorter in implementation. It is less error-prone and needs significantly less working memory in low Reynolds number flows. Our tests showed that the algorithm is less efficient in multiphase flows. To overcome this problem, further extension and the moments-only formulation was derived, inspired by the Multi-Relaxation Time (MRT) approach for single component multiphase flows.
\end{abstract}


\end{frontmatter}

\section{Introduction}
Computational fluid dynamics (CFD) is useful in many branches of science and technology, including those related to main civilization challenges of the utmost importance for the whole society e.g. weather forecast, climate, sport, medicine, oil recovery, and food industry \cite{Zajaczkowski11, Toparlar17, Xia02}. The most popular computational methods for CFD simulations are based on the direct discretization of the Navier-Stokes equations using appropriate numerical methods, e.g. finite differences, finite volumes, or finite elements \cite{Peiro05}. They are usually difficult to implement and require large computer resources as well as some tedious preprocessing of the input data (e.g. generation and storage of complex computational grids). In this context, the mesoscopic Lattice Boltzmann Method (LBM), based on the kinetic theory of gases, has recently been gaining more and more attention as a versatile and simple fluid solver that offers a wide range of potential applications \cite{Succi18}. 
One of the main limitations of the original LBM algorithm is a relatively high computer memory demand, as one has to store the distribution function for all fluid nodes. This limits the size of the samples that can be simulated in a single machine. Also, an increased number of memory accesses and complex memory access patterns in the propagation of distribution function may form a bottleneck for the parallel acceleration of the LBM \cite{Tomczak2019}. It was shown that in a GPU implementation the efficiency of the LBM solver saturates with the filling memory fraction \cite{Januszewski14}. Therefore, much research has been focused on improving memory efficiency of the LBM algorithm, including modifications of the main LBM algorithm \cite{Argentini04, Sheida2017} or data format and algorithms for sparse environments where most of the cells are getting fully blocked by obstacles \cite{Tomczak18, Valero2017}. 

In the standard LBM BGK algorithm the relaxation time $\tau$ may range from nearly 1/2 up to 1 (highly viscous flows), however, the choice $\tau=1$ is often made in the single relaxation time BGK approximation \cite{Wei12, Rao19, Khajepor19}. For example, $\tau=1$ was chosen in the gray LBM model used for porous media flows \cite{Chen08}, to compute the first predictor step and fictitious viscosity solution in the simulation of the mold filling process \cite{Szucki17}, in the LBM multicomponent flow simulation with comparison to Finite Volume Methods \cite{Shardt19} or in the immersed-boundary LBM for particles suspended in fluids \cite{Mountrakis17}. It was shown that the value of $\tau$ influences the accuracy of the LBM solver in flow through narrow pores \cite{Rao19}, and, $\tau=1$ case was in the best agreement with advanced multi-relaxation time schemes \cite{Pan2006}. This is also a special case for multiphase flows where EDM (Exact Difference Method) agrees well with the Shan-Chen method in terms of measured gas density error (which is minimum at $\tau=1$ for the Shan-Chen model) \cite{Brown14}. Moreover, it was reported as the best choice for the Shan-Chen two immiscible fluid simulation \cite{Hai00}. Also, it was shown that setting $\tau=1$ in BGK LBM provides optimal accuracy in time if solutions are compared to direct Navier-Stokes equations \cite{Worthing97}. Setting $\tau=1$ is also crucial for the fractional step formulation of LBM for high Reynolds flows \cite{Shu06}.

%
Here we fix the viscosity of the model and set the relaxation time to $\tau=1$ (from now We will use the codename LBMTau1) and modify the original LBM algorithm to a simpler, more compact and memory-efficient (an approach found i.e. in {\cite{Chen17, Zhou19}}). we provide a complete algorithm and first test it against Poiseuille flow with error scaling analysis. Then, we continue the with the  flow tests at varying Reynolds number and formulate criteria to calculate grid size necessary for stable simulations. 
We further develop the solver and use the multi relaxation time (MRT) version of the model at $\tau=1$ and compare its efficiency to the standard algorithm.
we show, that this approach leads to a significant memory drop and analyze this effect for various conditions and LBM models. 

\section{The Model}

The Lattice Boltzmann Method use the multi-dimensional velocity distribution function $f_k(\mathbf{x},t)$ to describe the state of the fluid. Function $f_k(\mathbf{x},t)$ corresponds to the probability that a molecule at position $\mathbf{x}$ at time $t$, is moving with velocity $\mathbf{e}_k$.
The original LBM algorithm consists of two steps: propagation and relaxation of the distribution function. It may be written as a discrete analogon to the Boltzmann transport equation (here with a linear approximation for the collision term) \cite{Guo13}:
\begin{equation} \label{lbm}
f_\mathrm{k}(\mathbf{x}+\mathbf{e}_\mathrm{k},t+1) = 
f_\mathrm{k}(\mathbf{x},t)-\frac{f_\mathrm{k}(\mathbf{x},t)-f^{eq}_\mathrm{k}(\mathbf{x},t)}{\tau},
\end{equation}
where k is the direction on the lattice, $f^{eq}$ is the equilibrium distribution function, $\mathbf{e}_\mathrm{k}$ is the lattice vector and $\tau$ is the relaxation time. By varying $\tau$, the kinematic viscosity of the fluid may be controlled \cite{Guo13}
\begin{equation}\label{viscosity}
v=c_s^2\left( \tau - 0.5 \right),
\end{equation}
where $c_s$ is the sound speed (dependent on the variant of the model, e.g. $c_s=1/\sqrt{3}$ for two dimensional D2Q9 model \cite{Guo13}). The equilibrium $f^{eq}$ is expressed in terms of macroscopic density $\varrho$ and velocity $\mathbf{u}$ of the flow field:
\begin{equation}\label{maxwellboltzmann}
f^{eq}_k=\omega_k\varrho\left(1+3\mathbf{e}_k\cdot\mathbf{u}+\frac{9}{2}(\mathbf{e}_k\cdot\mathbf{u})^2-\frac{3}{2}\mathbf{u}^2\right),
\end{equation}
where $\omega_k$ are direction weights \cite{Buick06}.
To include body force we may modify directly the momentum used for calculation of equilibrium (see e.g. \cite{Sukop07}).
The following sums over the
distribution function let us compute density and velocity:
\begin{equation}\label{rho1}
\varrho(\mathbf{x},t)=\sum_k f_k(\mathbf{x},t),
\end{equation}
\begin{equation}\label{u1}
\mathbf{u}(\mathbf{x},t)=\sum_k \mathbf{e}_k f_k(\mathbf{x},t).
\end{equation}

To save computer memory and eliminate the distribution function, we will first fix the relaxation time at $\tau=1$ \cite{Chen94, Zhou19}. With this assumption the transport equation (see Eq.~(\ref{lbm})) simplifies to:
\begin{equation} \label{LBMTau1}
f_\mathrm{k}(\mathbf{x}+\mathbf{e}_\mathrm{k},t+1) = f^{eq}_\mathrm{k}(\mathbf{x},t).
\end{equation}

Now, instead of keeping the values of $f_\mathrm{k}$ in memory, we plug Eq.~(\ref{LBMTau1}) into (\ref{rho1}) and (\ref{u1}). Thus, a new formulation will consist of computing macroscopic fields from the equilibrium distribution only
\begin{equation}\label{rhotau1}
\varrho(\mathbf{x},t)|_{\tau=1}=\sum_k f^{eq}_{ik}(\mathbf{x}+\mathbf{e}_\mathrm{k},t),
\end{equation}
\begin{equation}\label{utau1}
\mathbf{u}(\mathbf{x},t)|_{\tau=1}=\sum_k 
f^{eq}_{ik}(\mathbf{x}+\mathbf{e}_\mathrm{k},t)\cdot\mathbf{e}_k.
\end{equation}
where $f^{eq}$ is the Maxwell-Boltzmann distribution defined by Eq.~\ref{maxwellboltzmann}.
With the above two equations, we can write down an algorithm in which step by step the equilibrium distribution is computed from the macroscopic velocity and density and then use these values to make another iteration. In this way, the storage of $f_\mathrm{k}$ is eliminated.

For simplicity, we restrict our discussion to the D2Q9 model \cite{Guo13} (a two dimensional LBM model with nine lattice velocities $\mathbf{e}_\mathrm{k}$), where 
\begin{multline}
\mathbf{e}_\mathrm{k} \in \{ (0,0), (1,0), (0,1), 
(-1,0), (0,-1), (1,1), \\
(-1,1), (-1,-1),(1,-1)\},
\end{multline}
for $k=0\ldots8$ respectively.

To complete the picture, we need to account for the boundary conditions at the no-slip (zero tangent velocity) walls. 
\begin{figure}
\begin{center}
\includegraphics[width=0.95\columnwidth]{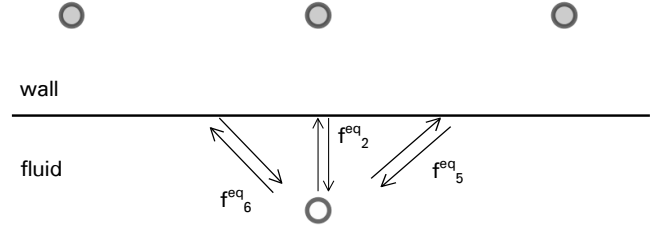}
\caption{\label{noslip0}Calculation of the equilibrium function next to the no-slip wall.}
\end{center}
\end{figure}
For fluid nodes located next to a no-slip wall, the normal components of the equilibrium distribution function must be reversed (see Fig.~\ref{noslip0}) and used in equations (\ref{rhotau1}) and (\ref{utau1}).
If we are at fluid node at $\mathbf{x}$ and the node $\mathbf{x}+\mathbf{e}_\mathrm{k}$ is of the no-slip type (a solid wall) we must use the following expression for $f^{eq}_\mathrm{ik}$ in Eqs.~(\ref{rhotau1}) and (\ref{utau1})
\begin{equation}\label{feqnoslip}
f_\mathrm{ik}^{eq}(\mathbf{x}+\mathbf{e}_\mathrm{k}) = f_\mathrm{k}^{eq}(\mathbf{x},\mathrm{u}=0, \mathrm{v}=0)=\omega_\mathrm{k}\cdot\varrho(\mathrm{x}).
\end{equation}
For example, for the wall located at the north, we need to reverse three populations that move towards the wall: $f^{eq}_6(\mathbf{x})$, $f^{eq}_2(\mathbf{x})$ and $f^{eq}_5(\mathbf{x})$ (see Fig.~\ref{noslip0}). Thus, for the north wall being no-slip we will have
%
\begin{multline}
\varrho_{\tau_1}(\mathbf{x},t) = 
f^{eq}_3(\mathbf{x}+\mathbf{e}_\mathrm{1})+
f^{eq}_1(\mathbf{x}+\mathbf{e}_\mathrm{3})+
f^{eq}_2(\mathbf{x}+\mathbf{e}_\mathrm{4})+\\
f^{eq}_6(\mathbf{x}+\mathbf{e}_\mathrm{8})+
f^{eq}_5(\mathbf{x}+\mathbf{e}_\mathrm{7})+
f^{eq}_6(\mathbf{x})+
f^{eq}_2(\mathbf{x})+
f^{eq}_5(\mathbf{x}).
\end{multline}
The first five terms on the right-hand side in the above equation are standard incoming populations from neighboring nodes, whereas the three last terms are the populations reflected from the northern wall and computed in-place at the node $\mathbf{x}$.
This procedure is used for all nodes adjacent to the walls. However, we do not need to write down an explicit expression for each orientation of the wall - it may be implemented by a simple expression in the algorithm. One has to check, if the neighboring node is a wall or not and choose Eq.~(\ref{maxwellboltzmann}) or (\ref{feqnoslip}) for equilibrium, accordingly.

\subsection{The LBMTau1 algorithm} 
\label{algorithm}

Using the procedure introduced in the previous section and summarized by Eqs.~(\ref{rhotau1}) and (\ref{utau1}), we formulate a complete algorithm for the LBMTau1 solver. Here $\mathrm{R}_{c}$, $\mathrm{U}_{c}$ and $\mathrm{V}_{c}$ are the macroscopic density, velocity (x) and velocity (y) fields at even (c=0) or odd (c=1) time steps. Thus, subscript $c$ is equal to 0 or 1 and denotes the grid number (we keep two copies of the grid to ping-pong data in the memory). Variables $i$, $j$ and $i_p$, $j_p$ are grid coordinates. Within the algorithm, we use the macroscopic density, velocity, lattice vector components $e_{k,x}$ and $e_{k,y}$ and the grid direction weights $\omega_{ik}$ to calculate the equilibrium distribution function $f^{eq}_{ik}$.

\begin{algorithm}
\begin{algorithmic}[5]
\STATE initialize flags and fields for fluid and solid nodes
\FORALL{fluid nodes (i, j)} 
 \STATE $\mathrm{U}_{c}(i, j) \gets 0$
 \STATE $\mathrm{V}_{c}(i, j) \gets 0$
 \STATE $\mathrm{R}_{c}(i, j) \gets 0$
\FORALL{directions k} 
\STATE $i_p \gets i + e_{k,x}$ (neighbour in the direction k)
\STATE $j_p \gets j + e_{k,y}$
\STATE $i_k \gets$ direction inverse to k
\IF{$(i_p, i_p)$ is fluid node}
 \STATE $r \gets \mathrm{R}_{1-c}(i_p,j_p)$
 \STATE $u \gets \left(\mathrm{U}_{1-c}(i_p,j_p)+f_x\right) / r$
 \STATE $v \gets \left(\mathrm{V}_{1-c}(i_p,j_p)+f_y\right) / r$
 \STATE $f^{eq}_{ik} \gets \omega_{ik}r \cdot$
 \STATE $(1-\frac{3}{2}(u^2+v^2)+3(e_{ik}^xu+e_{ik}^yv) + 
 \frac{9}{2}(e_{ik}^xu+e_{ik}^yv)^2
 )$
 \ELSE 
 \STATE $f^{eq}_{ik} \gets \omega_{k} \mathrm{R}_{1-c}(i, j)$
 \ENDIF
\ENDFOR
 \STATE $\mathrm{R}_{c}(i,j) \gets \mathrm{R}_{c}(i,j) + f^{eq}_{ik}$
 \STATE $\mathrm{U}_{c}(i,j) \gets \mathrm{U}_{c}(i,j) + e_{ik}^xf^{eq}_{ik}$
 \STATE $\mathrm{V}_{c}(i,j) \gets \mathrm{V}_{c}(i,j) + e_{ik}^yf^{eq}_{ik}$
\ENDFOR
\STATE Visualization (optional)
\end{algorithmic}
\caption{Complete time step of the LBMTau1 algorithm.}
\label{alg1}
\end{algorithm}

The algorithm starts with the initialization of macroscopic fields (line 1). At this point, we set up the flags for each node (flags denote if the node is occupied by fluid or solid). Also, the initial velocity and density fields are set up here (we start from zero velocity condition and density set to one). Next, we start the main loop over all fluid nodes (line 2) and for each of them compute the equilibrium distribution function from the local velocity and density. We also include the body force (lines 12-13 of the Algorithm \ref{alg1}). In the case of solid walls, we compute the equilibrium function by reflecting its normal components and assume zero velocity (no-slip boundary condition, line 17). Finally, we update the density and velocity by adding the populations that are incoming or being reflected from neighboring cells (lines 20-22). 
Implementation of this algorithm is straightforward - it contains $D+1$ tables (where $D$ is the dimension of the model), two loops, and one conditional (see the exemplary C/C++ implementation in \ref{appendixcode}).

\section{Validation and Results}

\begin{figure}
\centering
\includegraphics[width=0.95\columnwidth]{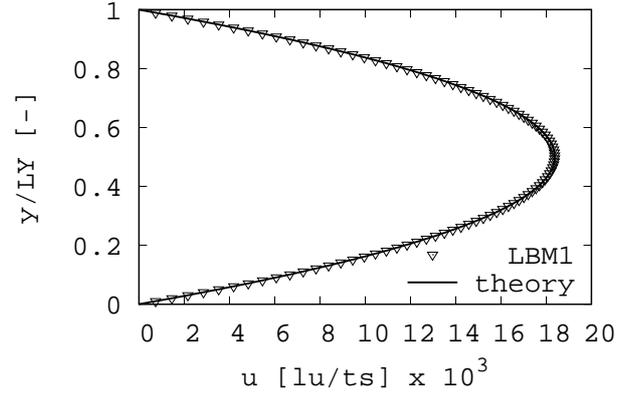}
\caption{The velocity profile in the Poiseuille Flow simulated using the LBMTau1 algorithm (points) compared to the analytical solution (solid line). The profile was taken along the y-axis perpendicular to the flow direction.\label{resultpoiseuille}}
\end{figure}

To verify the solver we run the steady-state flow in a straight rectangular two-dimensional channel and two-dimensional lid-driven cavity flows (standard tests performed in computational fluid dynamics). First, for the channel flow, we use the periodic conditions at the left and right system edges and the no-slip at the top and bottom walls. The external body force $f=2.5\cdot10^{-5}$ (lattice units) was used to generate a steady flow along the channel axis. we used a $200\times 100$ grid. To verify if the steady-state was reached we monitored the changes of velocity in the middle of the channel and used the convergence condition:
\begin{equation}
 \frac{\left|u_{n-1}-u_{n}\right|}{\left| u_n \right|} < \varepsilon,
\end{equation}
where $\varepsilon=10^{-7}$. The resulting velocity profile along the channel crossection is given in Fig.~\ref{resultpoiseuille}. we find an excellent agreement between the numerical and analytical solutions.
\begin{figure}
\includegraphics[width=0.95\columnwidth]{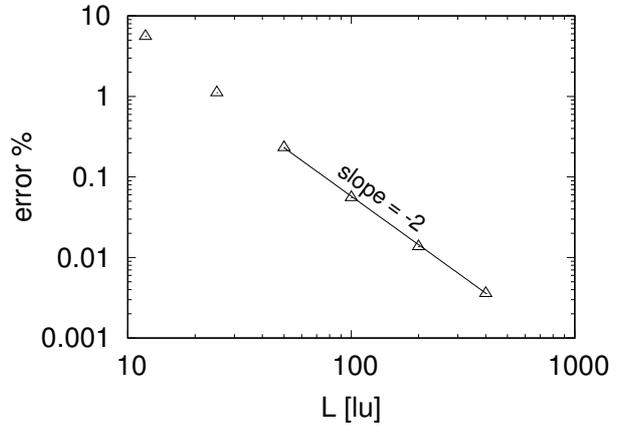}
\caption{Scaling of the relative error between LBMTau1 and analytical solution (Poiseuille Flow). The solid line represent best fit to $L^{-2}$ scaling taken at $L >= 40$.\label{resultpoiseuilleerror}}
\end{figure}
To quantify the agreement {we} repeat the simulations at varying grid size $L$ and calculate the percentage error of the solution 
%
	$e = 100 \cdot  \frac{\left|u-\tilde{u}\right|}{\left| \tilde{u} \right|}$,
%
where $\tilde{u}$ is the analytical value of velocity in the middle of the channel. we find that the error follows the power law and scales with the grid size as $L^{-2}$ (see Fig.~\ref{resultpoiseuilleerror}).

\begin{figure}
\begin{center}
\includegraphics[width=0.85\columnwidth]{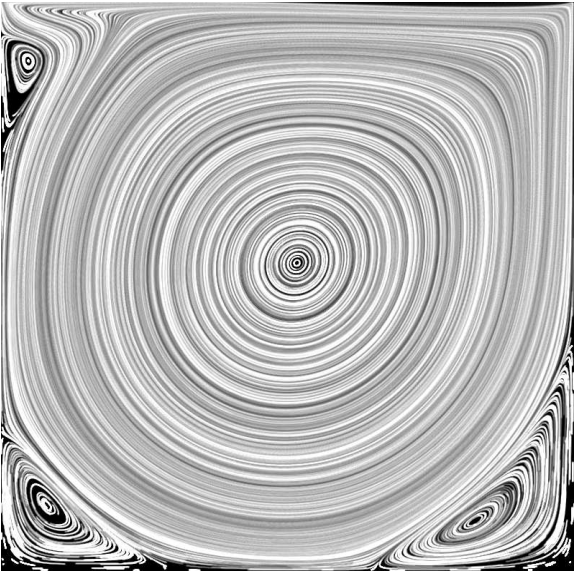}
\caption{The lid-driven cavity test run at Re=3200 calculated using the LBMTau1 code {(see \ref{appendixcode})} on a  $1100\times 1100$ grid. Visualization was made using massless tracers advected on top of the velocity field.\label{drivencavity}}
\end{center}
\end{figure}

Next, {we} run the standard lid-driven cavity flow in which the fluid is enclosed in a rectangular cavity with a top lid moving at a constant velocity \cite{Botella98}. The Dirichlet boundary condition $\mathbf{v}_\mathrm{lid}=(u_0,0)$ at the top boundary is applied. The no-slip condition is applied at the left, right, and bottom boundaries. 
\begin{figure}
\begin{center}
\includegraphics[width=0.99\columnwidth]{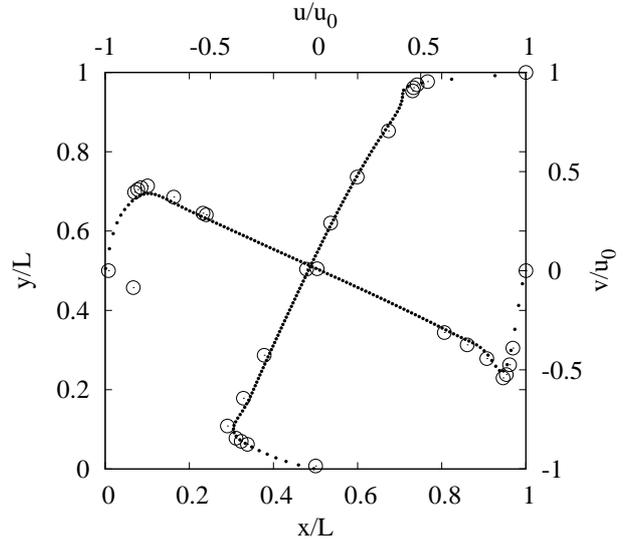}
\caption{Velocity profiles in lid-driven cavity at $Re=3200$. LBMTau1 results (solid dots) are compared to the literature benchmark data \cite{Ghia82} (open circles).\label{cavityprofiles}}
\end{center}
\end{figure}
{We} performed the simulation on a $1100\times 1100$ grid and velocity $u_0=0.4844$ at Re=3200. The simulation was continued as long as relative changes in the volumetric flux across vertical cross-section located at half of the system were larger than $1\%$. {We} checked the change between two timesteps at $\Delta t = 1000$ interval. In Fig.~\ref{drivencavity} we draw the streamlines on top of the final velocity field. The main vortex in the middle of the cavity, as well as vortex structures in corners of the cavity, are visible. The quantitative comparison with the multigrid method is given in Fig.~\ref{cavityprofiles}.

The final test of a time-dependent multiphase flow is the multiphase Shan-Chen model \cite{Shan93} in the LBMTau1 solver with the random initial conditions. 
\begin{figure}
\begin{center}
\includegraphics[width=0.96\columnwidth]{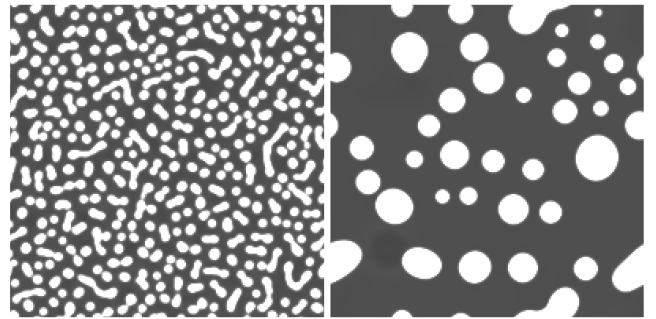}
\caption{LBMTau1 implementation of the multiphase simulation using the Shan-Chen model.}
\label{multiphase}
\end{center}
\end{figure}
The implementation of the Shan-Chen model was straightforward and we observed an expected phase separation effect (see Fig.~\ref{multiphase}). However, due to the two-fold loop over all neighbours of each computational node (once we need to go over neighbours and then over neighbours of each neighbour), we observed a significant drop in efficiency in the multiphase algorithm, if compared to standard LBM.

\section{Moments-only multirelaxation time LBMTau1}

The significant bottleneck of LBMTau1 in terms of performance is an unnecessary evaluation of $f^{eq}$ in each lattice node, which is mostly visible in the multiphase flows as found in the previous section. Let us look at evolution equation Eq.~(\ref{LBMTau1}) at $\tau=1$ (or $\omega=1$),
which could be rewritten as 
\[
f_{j}(\mathbf{x},t+\Delta t)=f_{j}^{eq}(\rho(\mathbf{x}-\mathbf{e}_{j},t),\mathbf{u}(\mathbf{x}-\mathbf{e}_{j},t)),
\]
to underline dependence on macroscopic variables. In the Multi-Relaxation Time (MRT) scheme, if we take the raw moments matrix $\mathbf{M}=M_{ij}$, and multiply
the evolution equation, we obtain
\[
M_{ij}f_{j}(\mathbf{x},t+\Delta t)=M_{ij}f_{j}^{eq}(\rho(\mathbf{x}-\mathbf{e}_{j},t),\mathbf{u}(\mathbf{x}-\mathbf{e}_{j},t)).
\]
Here, by definition the density is defined as:
\[
\rho=M_{1j}f_{j},
\]
whereas momentum reads:
\[
\rho\mathbf{u}=M_{kj}f_{j}\thinspace,k=2,3[,4].
\]
As a consequence, the evolution equation is reduced to:
\[
\rho(\mathbf{x},t+\Delta t)=M_{1j}f_{j}^{eq}(\rho(\mathbf{x}-\mathbf{e}_{j},t),\mathbf{u}(\mathbf{x}-\mathbf{e}_{j},t)),
\]
\[
\rho\mathbf{u}(\mathbf{x},t+\Delta t)=M_{kj}f_{j}^{eq}(\rho(\mathbf{x}-\mathbf{e}_{j},t),\mathbf{u}(\mathbf{x}-\mathbf{e}_{j},t))\thinspace,k=2,3[,4].
\]
Now we can derive closed formulas,
and skip explicit evaluation of $f^{eq}$ entirely. The main reason
for speedup in such a case is that we do not evaluate unnecessary degrees
of freedom of the system (higher-order moments) as they are relaxed
to equilibrium either-way. 
Nevertheless, the number of arithmetic operations
involved in the evaluation of density and momentum using those closed
statements could be significant. To optimize further, those statements
could be simplified using heuristic approaches and the computer algebra
system of choice. Statements for the evolution of D2Q9 lattice,
optimized by using PolyAlgebra package (part of TCLB software \cite{TCLB1}) are given
in supplementary material.

This approach could be used to optimize the Shan-Chen type multiphase
models. If we consider interaction potential force in form :
\begin{equation}
\mathbf{F}_{SC}=\psi(\mathbf{x})\sum_{i}\alpha_{i}\psi(\mathbf{x}+\mathbf{e}_{i})G(\mathbf{e}_{i}),\label{eq:f_sc}
\end{equation}
then for $\tau=1$, evolution equation could be once again written as:
as
\[
f_{j}(\mathbf{x},t+\Delta t)=f_{j}^{eq}(\rho(\mathbf{x}-\mathbf{e}_{j},t),\mathbf{u}(\mathbf{x}-\mathbf{e}_{j},t),\mathbf{F}_{SC}(\mathbf{x}-\mathbf{e}_{j},t)).
\]
which gives similar final result:
\[
\mathbf{M}f_{j}(\mathbf{x},t+\Delta t)=\mathbf{M}f_{j}^{eq}(\rho(\mathbf{x}-\mathbf{e}_{j},t),\mathbf{u}(\mathbf{x}-\mathbf{e}_{j},t),\mathbf{F}_{SC}(\mathbf{x}-\mathbf{e}_{j},t)).
\]

For a single component multiphase model, two approaches to evaluate
$\mathbf{F}_{SC}$ could be used.
In the first, one stores $\psi$ for latter use in neighbouring nodes in Eq.~(\ref{eq:f_sc}) loops. In the second, one recalculates $\psi$ of neighbouring nodes in-place anytime it is necessary. We found, that the first method significantly speeds up the algorithm at memory cost of additional scalar field. 


\subsection{Performance evaluation}

To evaluate the computational performance of presented models, we used
a model of fluid drop enclosed in a periodic domain. For an inter-particle force, we used the equation proposed by Kupershtokh \cite{kupershtokhEquationsStateLattice2009},
which provide superior stability compared to original Shan-Chen scheme remaining
identical from the implementation point of view, especially when
$\tau=1$. Details of both methods could be found in \cite{kupershtokhEquationsStateLattice2009,Shan93}.

A drop of liquid is placed in a periodic domain filled with vapor. We used $L\times L$ grids at $L$ from 128 to 8192. The final, largest grid used has filled almost completely the GPU memory of the NVIDIA GPU V100 card used for all tests. Thus, it was impossible to calculate final, largest grid using standard LBM (which clearly showed the advantage of LBMTau1 formulation). The simulation was run for several iterations until average velocity reached a steady state. 

\begin{figure}
\includegraphics[width=0.9\columnwidth]{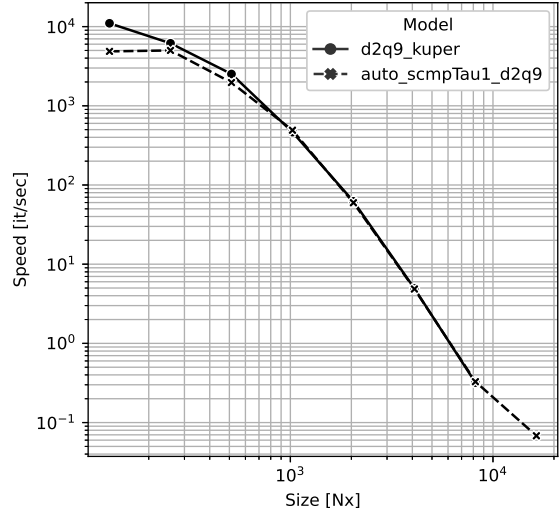}
\caption{Iteration speed expressed in iteration per second for two types of
solvers: LBMTau1 (crosses) and standard LBM implementation (filled circles). Both solvers were implemented in the TLCB framework \cite{TCLB1}. \label{fig:speed}}
\end{figure}

\begin{figure}
\includegraphics[width=0.9\columnwidth]{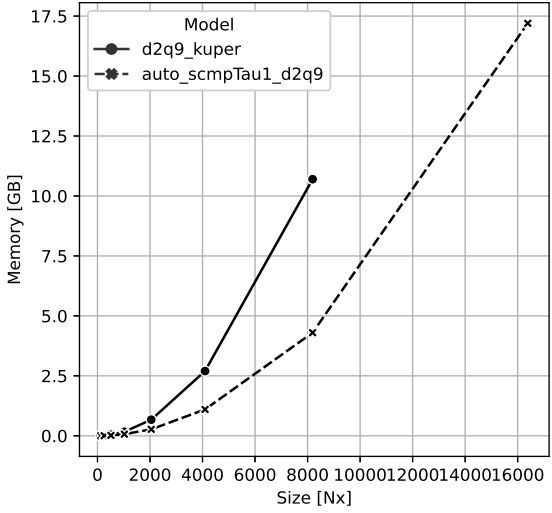}
\caption{Memory overhead for two types of solvers (see caption in Fig.~\ref{fig:speed} for reference). \label{fig:memory}}
\end{figure}

To quantify solver efficiency we plot the drop test iteration speed in function of the size of the system  (in lattice units) in Fig.~\ref{fig:speed}. We compare the two types of
solver \cite{TCLB1,Laniewski16}: classical "d2q9\_kuper" model where standard SCMP multiphase model is implemented (see \cite{TCLBMultiphase} for details) and the LBMTau1 variant
"auto\_scmpTau1\_d2q9" where no distribution function is used. 
Both simulations are carried out at the same viscosity and in the same two-dimensional square
domain. The memory usage of both solvers is compared in Fig.~\ref{fig:memory}.

\subsection{Convergence and accuracy of LBMTau1 and standard LBM}

The LBM method converges to the Navier-Stokes equation
in terms of
the small parameter $\epsilon$ used in the expansion, e.g. Chapman-Enskog
procedure. Two scalings are possible: acoustics one
$\epsilon=\Delta x=\Delta t$
and diffusive one $\epsilon=\Delta t=\Delta x^{2}$. To investigate
theoretical memory usage at preserved accuracy we restrict
ourselves to cases belonging
to the same viscous scaling series. For two viscosities, $\nu_{LB}$ and
$\nu_{LB1}=1/6$ the Reynolds number puts a restriction on
a both $\Delta x$ and $\Delta t$. For two grids, denoted by
LB (variable viscosity) and LB1 ($\tau=1$), it is required to
\[
\frac{U_{LB1}L_{LB1}}{\nu_{LB1}}=\frac{U_{LB}L_{LB}}{\nu_{LB}},
\]
where additionally diffusive scaling is defined as
\[
\epsilon=\Delta t=\Delta x^{2}\to0.
\]
We could now consider theoretical memory usage concerning
non-fixed viscosity LBM. We define $N_{LB}$ as grid resolution and $T_{LB}$ as the number of iterations (simulation time). By comparing Reynolds number for LB and LB1 cases, one gets
\[ \label{LBLB1}
\frac{T_{LB}N_{LB}^{2}}{T_{LB1}N_{LB1}^{2}}=6\nu_{LB}.
\]
On the other hand, from diffusive scaling time and spatial resolution are constrained by
\[
T_{LB}N_{LB}^{2}=\left(\Delta t\cdot\Delta x^{2}\right)^{-1}=\epsilon^{-2},
\]
which after substitution into Eg. \ref{LBLB1} gives
\[
\left(\frac{\epsilon_{LB1}}{\epsilon}\right)^{2}=6\nu_{LB}.
\]
From that we recover the spatial resolution ratio as
\[
\frac{N_{LB}}{N_{LB1}}=\frac{\Delta x_{LB1}}{\Delta
x_{LB}}=\sqrt{\frac{\epsilon_{LB1}}{\epsilon_{LB}}}=\left(6\nu_{LB}\right)^{1/4}.
\]
The ratio of the number of volume elements for LB case in relation to
LB1 is equal to
\[
N_{LB}^{D}=N_{LB1}^{D}\left(6\nu_{LB}\right)^{D/4},
\]
where $D$ denotes the number of spatial dimensions. If we now consider
floating-point variables for D2Q9 LB case (we do not consider constant factors in front of both expressions, e.g. double buffering, AA or SSS buffers):
\[
N_{LB}^{D}Q,
\]
and for D2Q9 LBMTau1 case
\[
N_{LB1}^{D}\left(D+1\right).
\]
Memory ratio is equal to
\[
\frac{N_{LB1}^{D}\left(D+1\right)}{N_{LB}^{D}Q}=\frac{D+1}{Q\left(6\nu_{LB}\right)^{D/4}}.
\]
For the D2Q9 model it is, thus, beneficial to have a larger grid
with $\tau=1$ as long as the second grid has a viscosity
\[
\frac{1}{54}<\nu_{LB}, (\tau>0.55).
\]
Similarly, for D3Q19 case, it is beneficial to have larger grids and $\tau=1$ as long as we compare with the standard algorithm at viscosity
\[
\frac{1}{6}\left(\frac{4}{19}\right)^{4/3}(\thickapprox0.021)<\nu_{LB},
(\tau>0.563).
\]
Those limits show that LBMTau1 is particularly efficient at low Reynolds number flows, where high spatial resolution is required, i.e. in porous media. 
The larger grid likely will require a larger
number of time steps (approximately square root of the spatial divisions ratio).
This drawback could be partially compensated
by faster iterations in LBMTau1. This, however is 
implementation, and model, dependent and rather hard to estimate theoretically.

\section{Discussion}

\begin{figure}
\begin{center}

\includegraphics[width=0.90\columnwidth]{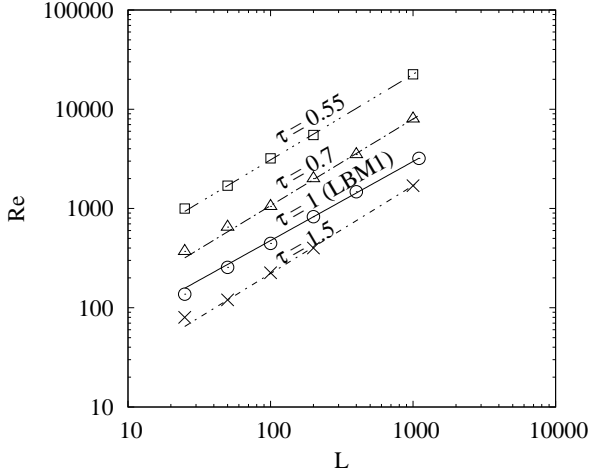}
\caption{
The maximum Reynolds number achieved at grid size L at various $\tau$. The data were obtained empirically by running several simulations for a given grid size. 
The least square fits (solid and dashed lines) to functions of the type $Re(L)=aL^{b}$ gave: 
a=3.7, b=0.89 for $\tau=1.5$, a=12, b=$0.8$ for $\tau=1$, a=19, b=$0.87$ for $\tau=0.7$ and  a=43, b=0.93 for $\tau=0.55$. 
\label{remax}}
\end{center}
\end{figure}

In this paper, a memory-saving algorithm for a simplified (fixed viscosity) LBM method is formulated and tested for flows with the relaxation time $\tau=1$. This results in an immediate relaxation of the local distribution function \cite{Chen94} and put some limitations on the range of parameters that may be used in the model. The Reynolds number is defined as
\begin{equation}\label{Reynolds}
Re=u_0L / \mu,
\end{equation}
where the viscosity $\mu=1/6$ (see Eq.~\ref{viscosity}). By changing $u_0$ or $L$ we control the Reynolds number which is now limited by the resolution of the grid only. To understand the limit we may estimate the Courant-Friedrichs-Lewy (CFL) condition. For velocity measured in the lattice units per time step we require, that $u_0$ (velocity of the top lid) fulfill $u_0 << 1$. Taking small $u_0$ stabilizes simulation, but at the same time slows down computation and require more memory as larger grids are required ($u_0$ and $L$ are the only parameters that may be changed in Eq.~\ref{Reynolds}). 
If we write the CFL condition as
\begin{equation}\label{U0}
u_0 = (Re/L) \cdot \mu << 1,
\end{equation}
and because $\mu=1/6$, to fulfill CFL criteria we should keep $Re/L << 6$. 
To check and validate this condition {we} run a series of simulations for the lid-driven cavity at $L$ from 50 to 1000 for increasing Reynolds number. If the simulation becomes unstable (and the solver crashed) after at least 1000 time steps, then the previous Re is taken as the maximum possible for the given lattice size. {We} repeated the procedure for various $\tau$ and collect the data in Fig.~\ref{remax}. The data for the smallest viscosity and $L<100$ agrees with \cite{Montessori14}. {We} notice that the lower relaxation time $\tau$ is, the higher Reynolds number may be achieved. However, for the relaxation time, $\tau=0.55$ the low resolution of the lattice leads to an inaccuracy in the solutions, especially in the regions where small vortices appear and in the center of the main vortex (data not shown). {We} found that if we keep $\tau=1$ then all converged solutions are of acceptable accuracy (see e.g. Fig.~\ref{cavityprofiles}). This finding agrees with the conclusions based on the linear stability theory where $\tau=1$ was suggested too \cite{Worthing97}. 
In practice, one could estimate the maximum Reynolds number using the grid size $L$ directly from the plot in Fig.~\ref{remax} or an empirical function fit to $Re(L)=aL^b$ given in the figure caption. 
The results for Re=3200 (see Fig.~\ref{cavityprofiles}) confirmed the stability of the solver for larger Reynolds numbers and large grids. There is only one outlier point for $u$ velocity component at $x\approx 0.46$ that has probably been a typo in the original tables provided in \cite{Ghia82}.

The main advantage of a new formulation is its relatively low memory consumption. For example, if we use the AB lattice access pattern in standard LBM (where an additional copy of the main lattice is kept in memory) the memory consumption is estimated from \cite{Sailfish19}
\begin{equation} \label{lbmmemory}
M_{\mathrm{LBM}} = (2Q+N_f)c,
\end{equation}
where Q represents lattice velocity directions (i.e. Q=9 in the standard D2Q9 model), $N_f$ is the number of macroscopic fields (density, velocity, etc.) and $c$ are bytes per single number ($c=4$ for float, $c=8$ for double-precision data). We may write that $N_f=D+1$, where D is the dimension of the model (D components of velocity plus density). Thus, in our case, if we eliminate the distribution function in the LBMTau1 algorithm, it will need only
\begin{equation} \label{LBMTau1memory}
M_{\mathrm{LBM}_1}=2N_f c
\end{equation}
bytes of the memory (the factor 2 appears because we store two copies of the macroscopic fields - one from the current and one from the previous time step). 
In D2Q9 model $N_f=3$ and $Q=9$. Thus, using equations (\ref{lbmmemory}) and (\ref{LBMTau1memory}) we have $M_{\mathrm{LBM}}=21c$ and $M_{\mathrm{LBM}_1}=6c$ respectively. This means the LBMTau1 algorithm needs $\Delta M \approx 76\%$ less memory than the original implementation. A similar calculation for the three-dimensional D3Q27 model gives $\Delta M\approx 86\%$. In practice, for the 2D $1000x1000$ lid-driven cavity flow in Fig.~\ref{drivencavity} we need $m=1000\cdot 1000 \cdot 84 = 84$ MB (megabytes) of memory in the standard LBM to store all simulation data. In LBMTau1, however, for the same grid size, we used only $m=1000\cdot 1000 \cdot 24=24$ MB. 
One should keep in mind, however, that in the basic LBMTau1 implementation this memory drop is true for low Reynolds number flows only (see Fig.~\ref{remax}) as higher Reynolds number may be achieved at smaller grids in the standard LBM. This problem, however, may be solved using i.e. fractional step approach for viscosity boost \cite{Shu06}, which we leave for future research.

The actual
memory gain measured from solver statistics for SCMP model (see Fig.~\ref{fig:memory}) is lower than in theoretical
discussions (theoretical $\approx44\%$, averaged $\approx 40\%$). This is due to the solver internal buffering designed for multi GPU communication.
Additionally, to speed up computations LBMTau1 variant of SCMP uses one additional global variable for inter-particle potential which sets a higher theoretical limit on memory gain. The theoretical limit could be lowered to $30\%$ at expense of additional computations and speed loss.

Apart from reduction of memory consumption, LBMTau1 could be optimized in terms of floating-point operations per lattice update as well. In the case of Shan-Chen type models, the proposed approach could outperform a classical model for the same parameter set. Such properties render such rewritten LBM a good candidate for low Re number flows with boundaries that could benefit from high grid resolution - for example, porous media flows.

We suggest that the LBMTau1 algorithm may provide a good starting point for fast and memory-efficient implementations of a solver in parallel environments, including graphics processors (GPUs), as the number of memory accesses decreases with decreasing memory demand of the main algorithm. 
However, to provide complete parallel implementation, one would need to consider memory access patterns used to compute macroscopic fields, which may not be the most efficient in the basic LBMTau1 implementation. To improve the parallel efficiency, we suggest using one of the improved memory layout algorithms and
data exchange algorithms used for the standard Lattice Boltzmann implementations. That includes an AA pattern in which one, instead of two buffers is used (thus, the two-factor reduction is achieved) and leads to $20\%$ performance gain compared to standard layout \cite{Bailey09}. 
Recently, the structure of arrays shifts and swap (SSS pattern) method based on the AA memory layout was also introduced \cite{Mohrhard19}. It comprises of an additional, separated array of directions for density function on the grid and has confirmed improved parallel efficiency as data access pattern is conserved between odd and even time steps \cite{Mohrhard19}.
The approach studied in this paper may be directly compared to the swap algorithm \cite{Mattila07}, where speedup is less than $1.5$ with a memory drop around 2 times if compared to the standard two-lattice approach. The main advantage of using LBMTau1 if compared to these schemes is the memory reduction achieved by removing the distribution function.

\section{Conclusions}

The presented LBMTau1 version of the LBM algorithm outperforms the original algorithm in terms of memory usage and is useful in large scale, low Reynolds number flows. This is important especially in systems where memory storage matters. This includes multiscale media e.g. porous and artery systems, where the flow at the microscale correlates with macroscopic properties of the medium.

Finally, it is rather surprising, how simple it is to implement a basic version of the LBMTau1 solver. The main function consists of a few lines of a simple C code (see \ref{appendixcode}). The ratio of the work needed to achieve useful results is relatively low, especially compared to any standard CFD solver. Thus, we believe, the solution provided in this paper may be also attractive in computational physics education. From a practical point of view, the LBMTau1 algorithm discussed here should be useful in applications where the original BGK Lattice Boltzmann was combined with the relaxation time $\tau=1$. For example, in \cite{Wei12, Khajepor19, Chen08, Szucki17, Mountrakis17, Brown14, Hai00, Worthing97, Shu06, Blaak00, Halliday13, Matyka08, Mendoza15, Shardt20} it is possible to save more than $75\%$ of the memory by using LBMTau1 described here.


\section{Acknowledgments}
We would like to thank Jonas Latt from the University of Geneva for discussions. Also, many remarks and first reading comments from Remigiusz Durka and Zbigniew Koza from the University of Wroc{\l}aw were very handful.

\appendix

\section{The LBMTau1 C code for lid-driven cavity} \label{appendixcode}


\definecolor{mGreen}{rgb}{0,0.6,0}
\definecolor{mGray}{rgb}{0.5,0.5,0.5}
\definecolor{mPurple}{rgb}{0.58,0,0.82}
\definecolor{backgroundColour}{rgb}{0.95,0.95,0.92}

\lstset{numbers=left, numberstyle=\tiny,language=C,frame=single,columns=fullflexible}


\begin{lstlisting}
float U[2][L][L], V[2][L][L], R[2][L][L]; 
int F[L][L];
const int ex[9]={0,1,0,-1,0,1,-1,-1,1};
const int ey[9]={0,0,1,0,-1,1,1,-1,-1};
const int inv[9]={0,3,4,1,2,7,8,5,6};
const float w[9]={4/9.,1/9.,1/9.,1/9.,1/9.,1/36.,1/36.,1/36.,
1/36.};
float U0=0.5;

void init()
{
 for(int i=0; i<L ; i++)  
 for(int j=0; j<L ; j++)  	  	
 {
  U[0][i][j]=V[0][i][j]=0;
  U[1][i][j]=V[1][i][j]=0;
  R[0][i][j]=R[1][i][j]=1;
  F[i][j]=0;
  
  if(j==0 or i==0 or i==L-1) F[i][j] = 1;
  if(j==L-1) U[0][i][j] = U[1][i][j] = U0;
 }
}

void LBMTau1(int c)
{
 float r,u,v,f;
 
 for(int i=0; i<L; i++)		
 for(int j=0; j<L-1; j++)
 if(F[i][j]==0)
 {	
  U[c][i][j]=V[c][i][j]=R[c][i][j]=0;

  for(int k=0; k<9; k++)
  {	
   int ip=i+ex[k], jp=j+ey[k], ik=inv[k];
 	
   if(F[ip][jp]==0)
   {
    r=R[1-c][ip][jp];
    u=U[1-c][ip][jp]/r;  
    v=V[1-c][ip][jp]/r; 
    
    f=w[ik]*r*(1-(3/2.)*(u*u+v*v)+3.*(ex[ik]*u+ey[ik]*v)
    +(9/2.)*(ex[ik]*u+ey[ik]*v)*(ex[ik]*u+ey[ik]*v));
   } 
   else
     f=w[ik]*R[1-c][i][j];

   R[c][i][j] += f;
   U[c][i][j] += ex[ik]*f;
   V[c][i][j] += ey[ik]*f;	  
  }	  
 }
}
\end{lstlisting}

\bibliographystyle{elsarticle-num-names}

\end{document}